\documentclass[lettersize,journal]{IEEEtran}
\usepackage{amsmath,amsfonts}
\usepackage{array}
\usepackage[caption=false,font=normalsize,labelfont=sf,textfont=sf]{subfig}
\usepackage{textcomp}
\usepackage{stfloats}
\usepackage{url}
\usepackage{verbatim}
\usepackage{graphicx}
\usepackage{cite}

\usepackage{multirow}
\usepackage{subfig}
\usepackage{float}
\usepackage{amsthm}
\theoremstyle{definition}
\newtheorem{defn}{Definition}

\usepackage{algorithmic}
\usepackage{algorithm}

\begin{document}

\title{XAV: A High-Performance Regular Expression Matching Engine for Packet Processing}

\author{Jincheng Zhong, Shuhui Chen, Chuan Yu
\thanks{This paper was produced by National University of Defense Technology. They are in Changssha, CN.}
}


\maketitle

\begin{abstract}

  Regular expression matching is the core function of various network security applications such as network intrusion detection systems. With the network bandwidth increases, it is a great challenge to implement regular expression matching for line rate packet processing. To this end, a novel scheme named XAV targeting high-performance regular expression matching is proposed in this paper. XAV first employs anchor DFA to tackle the state explosion problem of DFA. Then based on anchor DFA, two techniques including pre-filtering and regex decomposition are utilized to improve the average time complexity. Through implementing XAV with an FPGA-CPU architecture, comprehensive experiments show that a high matching throughput of up to 75 Gbps can be achieved for the large and complex \textit{Snort} rule-set. Compared to state-of-the-art software schemes, XAV achieves two orders of magnitude of performance improvement. While compared to state-of-the-art FPGA-based schemes, XAV achieves more than 2.5x performance improvement with the same hardware resource consumption.
\end{abstract}

\begin{IEEEkeywords}
Regular expression matching, finite state automaton, network intrusion detection, packet processing.
\end{IEEEkeywords}

\section{Introduction}
A regular expression (regex) comprises a sequence of characters, which describe a search pattern that can match a set of strings. For example, regex “[a-zA-Z0-9\_-]+@[a-zA-Z0-9\_-]+($\backslash$ .[a-zA-Z0-9\_-]+)+\$” describes a search pattern matching any email address. For their powerful expressiveness in defining search patterns, regexes are widely used in defining malicious features within many network security applications such as network intrusion detection systems (NIDS) and web application firewalls (WAF). Usually, regular expression matching (REM) is performed for every incoming packet in these network security applications. If one packet is found to match some regex, it indicates that the packet possibly contains some malicious pattern and further instructions will be performed accordingly.

In order not to affect business applications, network security applications must perform packet checking at line rate \cite{brunella2022hxdp, wang2022isolation, 2016A}. However, with the rapid growth of the Internet, the network bandwidth of data centers has rapidly increased to 40 Gbps or even 100 Gbps. To the best of our knowledge, very few REM schemes can achieve a matching throughput of tens of Gbps for large-scale regex rule-sets.

Targeting high matching throughput of tens of Gbps, this paper proposes a novel REM scheme -- XAV. XAV is named by its three matching stages: \textbf{X}or filter, \textbf{A}nchor DFA, and \textbf{V}erification. Finite state automaton is the basic way to implement regular expression matching, and nondeterministic finite automata (NFA) and deterministic finite automata (DFA) are two common forms of finite state automata. However, NFA and DFA both confront severe problems in implementing high-performance REM for large-scale rule-sets. The time complexity of the former is too high, while the space complexity of the latter is too large. 

This paper first puts forward the concept of anchor DFA, which is essentially a DFA compiled from regexes all starting with anchor “\^{}”. The application of anchor DFA greatly simplifies the regex semantics, thus avoiding the state explosion problem of DFA. Although anchor DFA has much lower space complexity than DFA, anchor DFA also introduces one big problem it needs to start one matching thread at every position of input to keep the semantics of original regexes. In order to make the average time complexity of anchor DFA close to that of DFA, two optimizations are proposed for anchor DFA. First, the xor filter, which is responsible for the matching of string parts in regexes, is used to reduce the number of anchor DFA matching threads. Second, some regexes are split into multiple parts to prevent a matching thread of the anchor DFA from matching too many input bytes. To cooperate with regex splitting, an extra verification stage is then introduced to perform the regex semantic verification for original regexes that have been split. Consequently, combining the two optimizations for anchor DFA, the three matching stages of our proposed REM scheme are xor filter, anchor DFA, and verification. 

We observe that the logic of the first two matching stages of XAV is simple and suitable for hardware implementation to achieve high matching performance. However, the logic of the last matching stage (verification) is complex and only suitable for software implementation. Therefore, an FPGA-CPU architecture is proposed to implement XAV, with implementing xor filter and anchor DFA in FPGA and implementing verification in CPU.

The main contributions of this paper are as follows:
\begin{enumerate}
    \item This paper first proposes to employ anchor DFA to solve the state explosion problem of DFA, thus achieving high-performance regex matching for large-scale rule-sets.
    \item Xor filter and an extra verification stage are introduced to greatly reduce the time complexity of anchor DFA. Combining the two stages with anchor DFA, a novel REM scheme XAV is proposed.
    \item A FPGA-CPU architecture is employed to implement XAV and detailed implementations are given in this paper. The compiler source code of XAV will be open-sourced once the paper is accepted.
    \item Comprehensive experiments are conducted to evaluate XAV. Experimental results show that XAV has great spatial and temporal performance and has achieved 75 Gbps throughput for the large \textit{Snort} rule-set.
\end{enumerate}

The rest of this paper is organized as follows. Section \ref{sec:related-work} introduces the related works and inspirations of the XAV scheme are elaborated in Section \ref{sec:inpirations-xav}. Next, the compilation procedure of XAV is given in Section \ref{sec:compilation-sav}, and the FPGA-CPU implementation of XAV is depicted in Section \ref{sec:scheme-implementation}. Finally, Section \ref{sec:evaluation} comprehensively evaluates the proposed scheme, and a conclusion is given in Section \ref{sec:conclusion}.

\section{Related works}
\label{sec:related-work}
There have been a large number of prior works concentrating on how to design an efficient REM engine over the past two decades. This paper divides the related works into two big categories: software REM schemes and hardware REM schemes. Table \ref{tab:related-work} provides a brief overview of related works.

\begin{table*}[!htbp]
    \label{tab:related-work}
    \begin{tabular}{ccccc}
    \hline
    Category                                                              & Scheme           & Constributions                                                                                                   & Limitations                                                                                           & Performance                                                                                                             \\ \hline
    \multirow{2}{*}{Software-based}                                       & HFA\cite{2007CuringHFA}/XFA\cite{2008DeflatingXFA}          & \begin{tabular}[c]{@{}c@{}}Proposes novel automata\\ to avoid the defects\\ of NFA and DFA\end{tabular}          & \begin{tabular}[c]{@{}c@{}}Only achieves a\\ tradeoff between\\ NFA and DFA\end{tabular}              & \begin{tabular}[c]{@{}c@{}}Hundreds of Mbps for medium-\\ scale rule-sets with\\ hundreds of regexes\end{tabular}       \\
                                                                          & Hyperscan\cite{wang2019hyperscan}        & \begin{tabular}[c]{@{}c@{}}Proposes regex decomposition\\ and prefiltering techniques\end{tabular}               & \begin{tabular}[c]{@{}c@{}}Performance seriously\\ degrades for\\ complex regexes\end{tabular}        & \begin{tabular}[c]{@{}c@{}}About 1 Gbps on large\\ Snort rule-set with\\ thousands of regexes\end{tabular}              \\
    \multirow{4}{*}{FPGA-based}                                           & Multi-stride NFA\cite{2012High} & \begin{tabular}[c]{@{}c@{}}Implements multi-stride\\ NFA using FPGAs\end{tabular}                                & Resource consuming                                                                                    & \begin{tabular}[c]{@{}c@{}}About 10 Gbps for medium-\\ scale rule-sets with\\ hundreds of regexes\end{tabular}          \\
                                                                          & Multi-stride DFA\cite{MDFA-yang2018high} & \begin{tabular}[c]{@{}c@{}}Implements multi-stride\\ DFA based on FPGA\end{tabular}                              & \begin{tabular}[c]{@{}c@{}}Suffers from DFA\\ explosion problem\end{tabular}                          & \begin{tabular}[c]{@{}c@{}}\textgreater{}100 Gbps for small\\ rule-sets with only\\ tens of regexes\end{tabular}        \\
                                                                          & FBDFA\cite{10.1093/comjnl/bxac138}            & \begin{tabular}[c]{@{}c@{}}Greatly alleviates the state\\ explosion problem of DFA\end{tabular}                  & Resource consuming                                                                                    & \begin{tabular}[c]{@{}c@{}}About 6 Gbps on large\\ Snort rule-set with\\ thousands of regexes\end{tabular}              \\
                                                                          & Pigasus\cite{Pigasus-zhao2020achieving}/Fidas\cite{Fidas-10.1145/3470496.3533043}    & \begin{tabular}[c]{@{}c@{}}Offloads network intrusion\\ detection to FPGA from\\ the system view\end{tabular}    & \begin{tabular}[c]{@{}c@{}}Only supports simple\\ regexes based on\\ string prefiltering\end{tabular} & \begin{tabular}[c]{@{}c@{}}80 Gbps for the Snort rule-\\ set with regexes containing\\ no string fragments\end{tabular} \\
    TCAM-based                                                            & Chain-based DFA\cite{2011Chain-TCAM}  & \begin{tabular}[c]{@{}c@{}}Employs TCAMs to perform\\ compressed-DFA matching\end{tabular}                       & \begin{tabular}[c]{@{}c@{}}Suffers from DFA\\ explosion problem\end{tabular}                          & \begin{tabular}[c]{@{}c@{}}Several Gbps for medium-\\ scale rule-sets with\\ tens of regexes\end{tabular}               \\
    \begin{tabular}[c]{@{}c@{}}Memory-centric\\ architecture\end{tabular} & Impala\cite{2020Impala}           & \begin{tabular}[c]{@{}c@{}}Designs automaton processing\\ architectures from the\\ transistor level\end{tabular} & \begin{tabular}[c]{@{}c@{}}Requires dedicated\\ hardware far\\ from the market\end{tabular}           & \begin{tabular}[c]{@{}c@{}}80 Gbps for large Snort\\ rule-set with thousands\\ of complex regexes\end{tabular}          \\ \hline
    \end{tabular}
    \end{table*}

\subsection{Software REM schemes}
Since DFA has the fantastical time complexity of O(1), a variety of studies \cite{kumar2022intelligent,kumar2021network,gong2022enabling, 2006AlgorithmsD2FA,2013ADFA,2015DesignRCDFA,2008AnDeltaFA} concentrate on trying to reduce the memory usage of DFA to support large-scale regex rule-sets. These studies have found that the state transition table of DFA is sparse and propose various compression methods accordingly. Although the proposed DFA compression techniques can cut down the memory usage of DFA by more than 95\%, linear memory compression cannot tackle the state explosion problem of DFA, which increases the memory footprint exponentially. Consequently, these REM schemes based on only DFA compression techniques can still only support small-scale rule-sets.

Another variety of studies realized that both NFA and DFA suffer from inherent defects. To achieve high-performance regex matching for large-scale regex rule-sets, novel automata such as HFA \cite{2007CuringHFA} and XFA \cite{2008DeflatingXFA} are designed to avoid the defects of NFA and DFA. Despite alleviating the high space complexity of DFA and the bad time complexity of NFA, these novel automata are essentially a tradeoff between NFA and DFA. These novel automata can only achieve medium matching performance for medium-scale rule-sets. With the regex rule-set size rising, the matching performance of these novel automata will severely drop. 

In recent years, regex decomposition methods \cite{haghighat2018hes,wang2019hyperscan,2019AMFA} become popular for achieving fast software REM. The key idea of regex decomposition methods is to extract string parts from original regexes first. Then taking fast string matching as the entry to regex matching, these REM schemes try to avoid expensive automata processing. Since string matching can be two magnitudes faster than regex matching \cite{wang2019hyperscan} and string parts in regexes are rarely matched to trigger further regex matching, regex decomposition methods have achieved great matching performance. Hyperscan \cite{wang2019hyperscan}, the most famous software REM scheme, claims it can achieve 1 Gbps throughput on \textit{Snort} rule-set with one CPU core.

\subsection{Hardware REM schemes}
\subsubsection{FPGA-based schemes}
A large number of studies utilize FPGA to accelerate regex matching \cite{10.1145/3314576,2017REAPR,2007Optimization-FPGA,2008Compact-FPGA,2012High}. Because of the inherent massive parallel logic, FPGA is an ideal platform to perform regex matching based on NFA processing. NFA can contain multiple active states at a time, and multiple state transitions need to be performed to consume one byte of input. Therefore, the software is slow for NFA processing. However, FPGA can perform the multiple state transitions of NFA in parallel. As a consequence, the time complexity of NFA in FPGA becomes O(1), and in each cycle, one byte of input can be handled. To achieve higher matching throughput, \cite{2010A-FPGA,2008High-FPGA,2012High} further propose multi-stride NFA to handle multiple characters each cycle on FPGA. Benefiting from the multi-stride technique, the FPGA NFA scheme \cite{2012High} even achieves a high matching throughput of 10 Gbps for \textit{Snort} rule-set. Although FPGA NFA schemes can achieve high matching performance for complex rule-sets, rule-set updates in FPGA NFA schemes are extremely tricky because FPGA circuit synthesis is required to map NFA state to FPGA circuits.

Some FPGA REM schemes \cite{2013A-FPGA-DFA,2014A-FPGA-DFA,2018A-FPGA-DFA,10.1093/comjnl/bxac138} also employ DFA to perform regex matching. The biggest advantage of employing DFA instead of NFA in FPGA is that the rule-set update for FPGA DFA schemes only needs to rewrite the FPGA memories, which is a fast procedure. However, DFA confronts the state explosion problem. Although in \cite{10.1093/comjnl/bxac138}, a regex decomposition method is used to tackle the state explosion problem, the space complexity of DFA is still too high for FPGA implementation. In the FPGA DFA scheme proposed in this paper, the space complexity is further reduced significantly. Consequently, our proposed novel scheme can achieve much better matching performance with close hardware resources.

In recent years, there have been researches \cite{Pigasus-zhao2020achieving, Fidas-10.1145/3470496.3533043} into accelerating network intrusion detection based on FPGAs from the system view. Zhao et al. \cite{Pigasus-zhao2020achieving} design the Pigasus IDS employing an FPGA-CPU architecture. Pigasus has implemented the majority of Snort functionalities (including TCP reassembly, fast string pattern matching, and packet header matching) on the FPGA. Packets that pass the fast string pattern matching and header matching are sent to the CPU for regex matching. Pigasus claims that it can achieve 100 Gbps packet processing throughput with one server and an FPGA board. However, Pigasus focuses on accelerating Snort NIDS from a system view and only employs some other available regex matching engine. 

Alibaba Group proposes Fidas \cite{Fidas-10.1145/3470496.3533043}, which is an FPGA-based intrusion detection offload system and has been deployed in their production data center. Unlike Pigasus, Fidas fully offloads the primary NIC, rule pattern matching, and traffic rate classification with no CPU overhead. Fidas uses a multi-level filter-based approach for efficient regex processing. It is reported that Fidas can achieve about 80 Gbps packet processing throughput in the paper. Although these system works like Pigasus and Fidas remarkably reduce CPU overhead by offloading packet processing to FPGA, they only support simple regex rules with many string fragments. As network threats become more complex and diverse, network intrusion detection gradually needs to support more and more complex regex rules. This paper proposes XAV trying to achieve high-performance regex matching for large-scale and complex rule-sets.

\subsubsection{TCAM-based schemes}
Some REM schemes \cite{2010CompactDFA-TCAM,2010Fast-TCAM,2011Chain-TCAM} employ TCAMs to perform regex matching. TCAMs are very effective for storing the state transition table of DFA. Employing TCAM to perform DFA matching only needs one TCAM query for handling each byte of input. However, TCAM-based REM schemes confront a similar problem with DFA compression methods: TCAMs can not help DFA tackle the state explosion problem. What is worse, although TCAM-based REM schemes can easily achieve a high matching throughput, they are very power-hungry for each TCAM query is a brute-force search of every TCAM memory entry.

\subsubsection{Memory-centric architecture schemes}
In recent years, several memory-centric architecture schemes \cite{2014An-AP,2017Cache-CA,2020Impala} are proposed to improve the performance of automata processing. Different from employing existing hardware like FPGA and TCAM, these memory-centric architecture schemes redesign hardware to fit regex matching. The Micron Automata Processor (AP) \cite{2014An-AP} provides a DRAM-based dedicated automata processing chip, while CA \cite{2017Cache-CA} and Impala \cite{2020Impala} modify the structure of the last-level cache SRAM for automata processing. Despite up to 80 Gbps throughput being claimed by Impala, the dedicated hardware design is far from the market.

\section{Motivations for XAV}
\label{sec:inpirations-xav}
\subsection{Anchor DFA}
\label{sec:motivation-anchor-dfa}
As a popular way to perform regex matching, DFA suffers from the state explosion problem \footnote{The famous state explosion problem of DFA is introduced in Appendix \ref{sec:preliminary}.} and requires impossibly large memory for supporting large-scale rule-sets. However, it is found that when a regex set is compiled into an anchor DFA (see definition \ref{defn:anchor-dfa}), the state explosion tends not to happen, and the size of the anchor DFA is small. It is reasonable that the semantics of one regex starting with anchor "\^{}" is much simpler than the same regex starting with ".*" \footnote{Note that the starting ".*" of one regex is usually omitted in practical rule-sets.}. The simpler the semantics of a regex set, the smaller the number of DFA states required to represent the semantics. Therefore, the anchor DFA compiled from all regexes starting with anchor "\^{}" is much smaller than the DFA compiled from all the same regexes starting with ".*".

\begin{defn}[anchor DFA]
\label{defn:anchor-dfa}
An anchor DFA refers to a DFA compiled from regexes all starting with anchor "\^{}". To compile a set of regexes into an anchor DFA, each regex is added with one anchor "\^{}" to the head first.
\end{defn}

Experiments are carried out to show that anchor DFA can significantly reduce the number of states. Four rule-sets from Regex \cite{Regexworkload_becchi2008workload} benchmark suit are used for the experiment, and each of them is compiled into a DFA and an anchor DFA, respectively. The size comparison of the DFA and the anchor DFA on each rule-set is shown in Table \ref{tab:anchorDFAsize-comparison}. For each of the four rule-sets, the anchor DFA size is much smaller than the DFA size. For each of the last three rule-sets, when all regexes in the rule-set are compiled into a DFA, state explosion occurs and the DFA construction is far from completion even when the state number exceeds one million. However, while compiling all regexes in each of the three rule-sets into an anchor DFA, DFA construction is finished when the size reaches only tens of thousands. From the above experiments, it is obvious that anchor DFA can effectively alleviate the state explosion problem of DFA.

\begin{table}[tb]
\centering
\caption{Comparison of the DFA size and the anchor DFA size on different rule-sets}
\label{tab:anchorDFAsize-comparison}
\begin{tabular}{ccc}
\hline
Rule-set             & DFA                & Anchor DFA \\ \hline
bro217.re    & 6533               & 2155       \\ 
ranges1.conf\_300-0.re   & \textgreater{}1000000 & 11509      \\ 
ranges05.conf\_300-0.re  & \textgreater{}1000000 & 11639      \\ 
dotstar0.3.conf\_300-0.re & \textgreater{}1000000 & 43679      \\ \hline
\end{tabular}
\end{table}

\begin{figure}[tb]
    \centering
    \centerline{\includegraphics[width=0.35\textwidth,trim=0 350 580 11,clip]{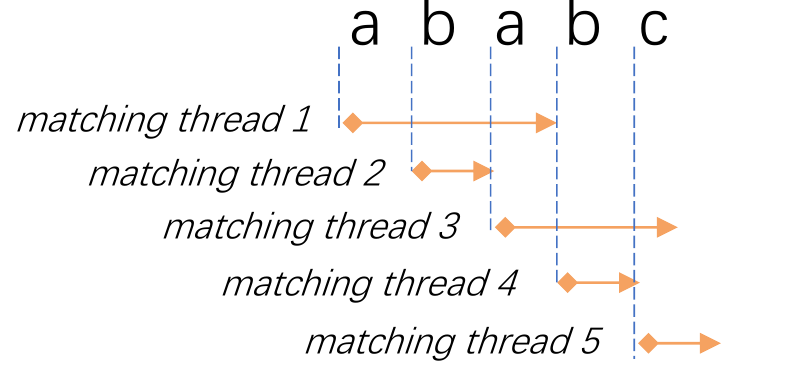}}
    \caption{Example of Anchor DFA matching procedure.}
    \label{fig:matching-thread}
\end{figure}

Although anchor DFA can greatly reduce the number of DFA states, the semantics of each regex has changed in anchor DFA for that an anchor "\^{}" is added to the head. To maintain the semantics of original regexes starting with ".*", anchor DFA should start a matching thread (see definition \ref{defn:matching-thread}) at each position of the input. For example, if $RE_1$="abc" is compiled into an anchor DFA $A_1$, now $A_1$ can only match "\^{}abc". For the input text $T_1$="ababc", to find the occurrences of $RE_1$ in $T_1$, $A_1$ should start a new matching thread at each position of $T_1$ as shown in Fig.\ref{fig:matching-thread}. And can only matching thread 3 starting at position 3 find the occurrence of $RE_1$. If only one matching thread (i.e., matching thread 1 in the figure) is started at position 1, which is the same as the DFA matching scheme, then no occurrence of $RE_1$ will be found.

\begin{defn}[matching thread]
\label{defn:matching-thread}
For an anchor DFA, a matching thread refers to the entire DFA matching process, which starts from the initial state and ends when arriving at the dead/trap state or the end of the input.
\end{defn}

Because each matching thread can require a number of state transitions, starting a new matching thread at each position of input indicates that anchor DFA has extremely high time complexity. Therefore, directly applying anchor DFA to regex matching is not an advisable option. However, some optimizations can greatly reduce the time complexity of anchor DFA.

\subsection{Optimizations to Anchor DFA}
In this section, two optimizations are proposed to reduce the time complexity of anchor DFA. The first is to employ one pre-filter to prevent anchor DFA from starting new matching threads in most positions. The second is to limit the access depth (see definition \ref{defn:access-depth}) of each matching thread through regex decomposition.

\begin{defn}[access depth]
\label{defn:access-depth}
The access depth of one matching thread refers to the total number of state transitions needing to be performed by the matching thread. 
\end{defn}

\subsubsection{Pre-filter}
In most practical DPI-based network security applications, input traffic rarely matches any rule in the rule-set \cite{2019A,becchi2007hybrid,xu2016survey}. Further, it has been observed that one pre-filter built from string fragments of each rule can be very efficient \cite{wang2019hyperscan,haghighat2018hes,haghighat2018toward,choi2016dfc}. Therefore, using the string fragments of each regex, one pre-filter can be constructed to prune most anchor DFA matching threads.

Taking $RE_2$="ab.*cd" as an example, while $RE_2$ is compiled into anchor DFA $A_2$, the string fragment $S$="ab" can be used to construct a pre-filter of $A_2$. Assume that the input text is $T_2$="aabcd". As $S$ is only matched by the pre-filter starting at position 2 of $T_2$, only one matching thread of $A_2$ needs to be started at position 2 in the whole matching process. For all other positions that are not matched by the pre-filter, no matching thread of $A_2$ is required. 

A number of algorithms for multi-string pattern matching, such as \cite{aho1975efficient,norton2004optimizing,wu1992fast,moraru2012exact,stylianopoulos2020multiple}, can be selected as the pre-filter. In this paper, the xor filter newly proposed in \cite{2020Xor} is used for that the xor filter is fast and small. What is more, the xor filter is friendly for hardware implementation. Therefore, implementing xor filters in FPGAs can help XAV achieve a high matching throughput.

\subsubsection{regex decomposition}
\label{sec:motivation-regex-decomposition}

Although the introduction of a pre-filter can significantly reduce the number of anchor DFA matching threads, the access depth of some unpruned matching threads may be very large, thereby severely downgrading the matching performance. 

Taking the example above, when $S$ is matched by the pre-filter, as ".*" (i.e., dot-star) in $RE_2$ means to match any character any time, the activated matching thread of $A_2$ will run all the time until one match occurs or the input ends. Besides the dot-star components in one regex, all kinds of long regex components (see definition \ref{defn:long-component}) can possibly lead to a large access depth for one anchor DFA matching thread.

\begin{defn}[long regex component]
\label{defn:long-component}
One long regex component refers to a regex component that can cause a large expected access depth for one matching thread. Note that the minimum expected access depth for a long regex component can be flexibly defined in the specific implementation. Generally, long regex components include dot-star, almost dot-star such as "[\^{}\textbackslash n]+", and big character classes with a large counting constraint such as "[\^{}\textbackslash n]\{500\}".
\end{defn}

To limit the expected access depth of each activated matching thread, each regex can be decomposed into lsREs (see definition \ref{defn:lsRE}) and lusREs (see definition \ref{defn:lusRE}). Only the lsREs of each regex are employed to construct the anchor DFA, while the matching of lusREs is left to a further verification stage. In this way, the access depth of each anchor DFA matching thread is limited to a small value, thereby reducing the time complexity of anchor DFA.

\begin{defn}[lsRE]
\label{defn:lsRE}
One lsRE \footnote{LsRE is an abbreviation of \textbf{L}ength re\textbf{S}tricted \textbf{RE}gex, which means that the expected length of each matching text of lsRE is small.} refers to a regex fragment with no long regex component. For example, "abc" and "user=[0-9]\{32\}" are both lsREs.
\end{defn}

\begin{defn}[lusRE]
\label{defn:lusRE}
One lusRE \footnote{LusRE is an abbreviation of \textbf{L}ength \textbf{U}nre\textbf{S}tricted \textbf{RE}gex, which means that the expected length of each matching text of lusRE can be very large.} refers to a regex fragment that contains at least one long regex component. For example, "ab.*cd" and "ab[\^{}\textbackslash n]\{500\}" are both lusREs. 
\end{defn}

For example, $RE_2=$"ab.*cd" can be split into two lsREs, $lsRE_1$="ab" and $lsRE_2$="cd", and one lusRE, $lusRE_1$=".*". We complile $lsRE_1$ and $lsRE_2$ into one anchor DFA $A_3$. Now, the maximum access depth of each matching thread of $A_3$ is limited to 3, which is the length of $lsRE_1$ or $lsRE_2$ plus one. As for the matching of $lusRE_1$, only when $lsRE_1$ and $lsRE_2$ are sequentially matched by $A_3$, will the further verification stage perform it.

The details of how to achieve the regex decomposition and further verification will be described in Section 
\ref{sec:compilation-sav}.

\subsection{XAV}
\begin{figure}[tb]
    \centering
    \centerline{\includegraphics[width=0.5\textwidth,trim=0 655 410 0,clip]{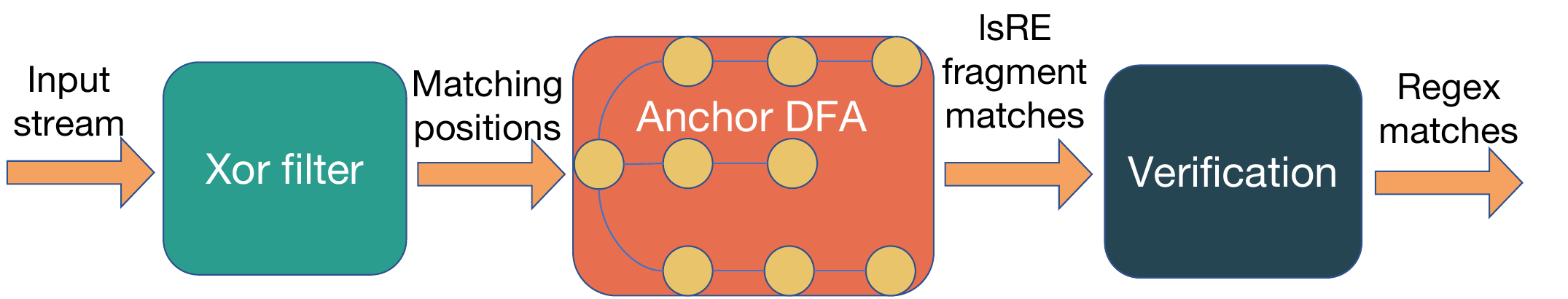}}
    \caption{The architecture of XAV.}
    \label{fig:SAV-arch}
\end{figure}

Combining the above two optimizations with anchor DFA, this paper proposes a novel REM scheme named XAV, which represents the three matching stages of the scheme, \textbf{X}or filter, \textbf{A}nchor DFA and \textbf{V}erification. Anchor DFA tackles the state explosion problem to support large-scale rule-sets, while the xor filter and the verification stage reduce the time complexity to achieve high matching performance. The architecture of XAV is as shown in Fig.\ref{fig:SAV-arch}.

Firstly, the input is matched by the xor filter, where matches occur only at a small proportion of input positions. Then, these suspicious positions are transmitted to the anchor DFA engine, which will start a new matching thread at each of them. The anchor DFA will find all occurrences of each lsRE fragment and report them to the further verification engine. In the verification engine, the matching results of each lsRE fragment are recorded and verified if any original regex matches the input.

To clarify the overall matching procedure of XAV, a simple example is given below. For the rule-set with one regex, $RE_2$="ab.*cd", and the input $T_2$="aabcd", the xor filter will report that the input matches at positions 2 and 4. Through two matching threads starting at positions 2 and 4, the anchor DFA can find that $lsRE_1$="ab" matches at position 2 and $lsRE_2$="cd" matches at position 4. When the verification engine receives that $lsRE_2$="cd" matches at position 4, since it knows that $lsRE_1$ has been matched at position 2, it can determine that $RE_2$ matches the input.

For the use of anchor DFA, the state explosion problem can be avoided when compiling large-scale rule-sets. Therefore, XAV has a low space complexity and can support the parallel matching of hundreds to thousands of regexes. Besides, benefitted from the two optimizations to anchor DFA, the number of activated anchor DFA matching threads is small, and the expected access depth of each matching thread is limited. Consequently, XAV has a low matching time complexity and can deliver a high matching performance. The detailed implementation of the XAV scheme is described in the following two sections.

\section{Compilation of XAV}
\label{sec:compilation-sav}

This section elaborates on the compilation procedure of XAV, which mainly includes regex decomposition, xor filter building, anchor DFA construction, and the verification engine building.

\begin{figure}[tb]
    \centering
    \centerline{\includegraphics[width=0.5\textwidth,trim=0 470 630 0,clip]{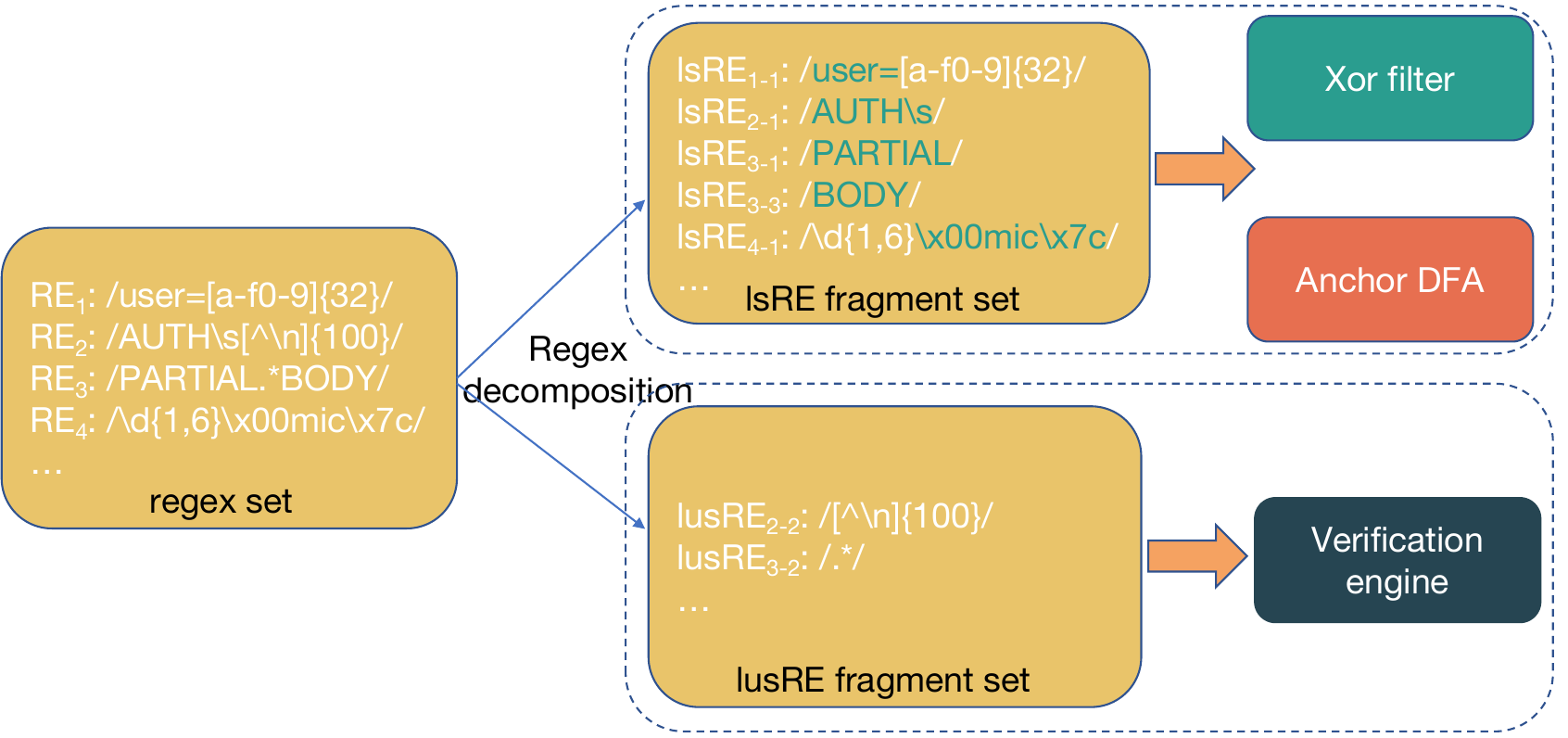}}
    \caption{The compilation procedure of XAV.}
    \label{fig:compilation-overview}
\end{figure}

The overall compilation procedure is shown in Fig.\ref{fig:compilation-overview}. Firstly, each regex is split into lsRE fragments and lusRE fragments through the regex decomposition. Then the xor filter and anchor DFA are constructed from the lsRE fragment set. While the lusRE fragment set is employed to build the verification engine.

\subsection{Regex decomposition}
To implement regex decomposition, each regex is first parsed to a component tree, which will be introduced below. By identifying long regex components in the component tree, the regex is then split into lsRE fragments and lusRE fragments at the position of each long regex component.

\subsubsection{Regex parsing}
According to regex semantics, four component classes are defined in this paper, which are character class components, concatenation components, alternation components, and repetition components, respectively. Given regex components $R$ and $S$, the definitions of the above four component classes are listed below:

\begin{itemize}
    \item{\textbf{character class component:} A character class component refers to the most basic regex component that matches a set of literal characters. Character class components include single literal characters (such as "a") and character classes (such as "[0-9]").}
    \item{\textbf{concatenation component ($RS$): } A concatenation component refers to a set of regex components that are concatenated one by one. To successfully match the concatenation component, one text needs to match the set of regex components sequentially. For the concatenation component of $R$ and $S$, each accepted text of ($RS$) should match $R$ and $S$ in order.}
    \item{\textbf{alternation component ($R|S$): } An alternation component refers to the set union of a regex component set. If one text matches any one regex component in the regex component set, the text matches the alternation component. For the alternation component of $R$ and $S$, each accepted text of ($R|S$) matches either $R$ or $S$. }
    \item{\textbf{repetition component ($R\{m,n\}$): }} A repetition component refers to a component with counting constraints. For the repetition component of $R$, with counting constraints $\{m,n\}$ ($0\le m\le n$ and $n$ can be infinite), each accepted text of $R\{m,n\}$ should match $R$ consecutively for $m$ to $n$ times.
    \end{itemize}

\begin{figure}[tb]
    \centering
    \centerline{\includegraphics[width=0.4\textwidth,trim=0 570 880 0,clip]{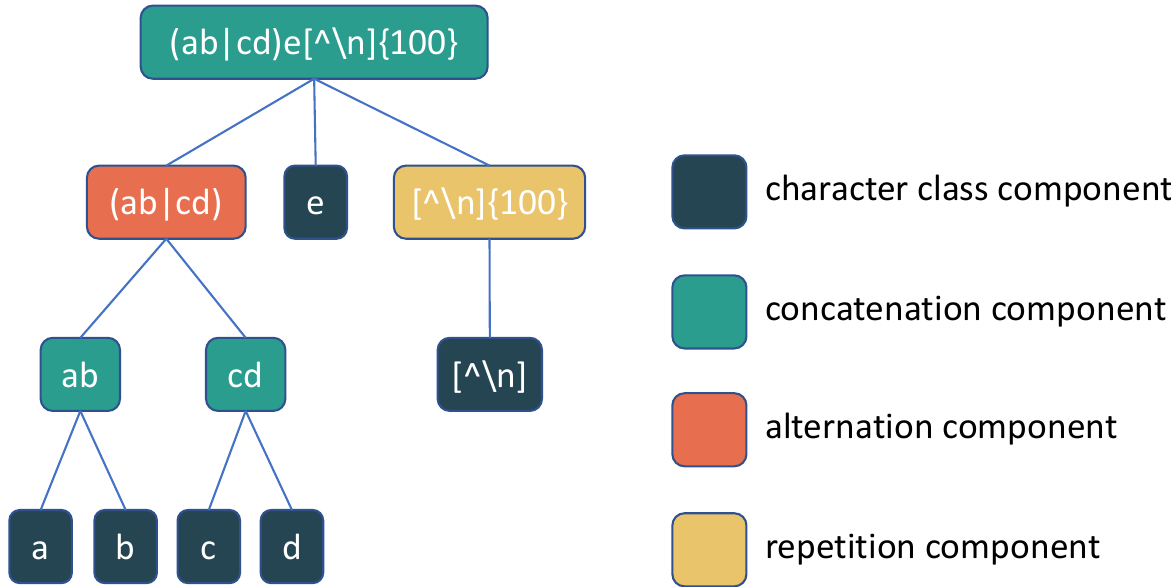}}
    \caption{The component tree for $RE_3$.}
    \label{fig:component-tree}
\end{figure}

Based on the concepts of the four regex components, each regex can be parsed into a component tree. For example, the regex component tree parsed from regex $RE_3$="(ab$|$cd)e[\^{}\textbackslash n]\{100\}" is as shown in Fig.\ref{fig:component-tree}.

\subsubsection{Regex splitting}
\label{sec:regex-splitting}
Based on the component tree of one regex, it is easy to identify where one long regex component is and therefore to split the regex. The definition of long regex components has been given in definition \ref{defn:long-component}. In our actual implementation, one character class component matching more than 128 out of 256 characters with a counting constraint larger than 50 is regarded as a long regex component.

To split one regex into lsRE fragments and lusRE fragments, the corresponding component tree is first traversed to find all the positions of long regex components. Then, the regex is divided into lsRE fragments and lusRE fragments at the position of each long regex component. Finally, the unfriendly lsRE fragments will be merged into neighbor lusRE fragments. Unfriendly lsRE fragments refer to those lsRE fragments that can not adapt to the xor filter, which will be discussed in the remainder of this paper.

Taking $RE_3$="(ab$|$cd)e[\^{}\textbackslash n]\{100\}" as an example, the repetition component "[\^{}\textbackslash n]\{100\}" will be identified as a long regex component. Then, according to the position of the long regex component, $RE_3$ is split into one lsRE fragment (i.e., "(ab$|$cd)e") and one lusRE fragment (i.e., "[\^{}\textbackslash n]\{100\}").

\subsection{Xor filter construction}
After decomposing each regex in the rule-set, one lsRE fragment set and one lusRE fragment set are generated. The lsRE fragment set is employed to construct the xor filter and the anchor DFA, while the lusRE fragment set is used for the building of the verification engine.

As xor filter only supports the matching of ldREs (see definition \ref{defn:ldRE}). For building the xor filter, given a length limitation \footnote{The length limitation of each ldRE, $L_T$, is set to 8 in detailed implementation.}, the ldRE fragment with the highest pre-filtering effectiveness is first extracted from each lsRE fragment. 

\begin{defn}[ldRE]
\label{defn:ldRE}
One ldRE \footnote{LdRE is an abbreviation of \textbf{L}ength \textbf{D}etermined \textbf{RE}gex, which means that each accepted text of one ldRE has the same length.} refers to a regex fragment that comprises only character class components. Note that some repetition components, of which the sub-component is a character class component and the counting constraint is a fixed number, can be regarded as multiple character class components. For instance, "abc", "ab[a-z]", "ab[a-z]\{3\}" are all ldREs.
\end{defn}

Taking the lsRE fragment set in Fig.\ref{fig:compilation-overview} as an example, ldREs "user=", "AUTH\textbackslash s", "PARTIAL", "BODY", and "\textbackslash x00mic\textbackslash x7c" will be extracted from $lsRE_{1-1}$, $lsRE_{2-1}$, $lsRE_{3-1}$, $lsRE_{3-3}$ and $lsRE_{4-1}$ in the compilation procedure, respectively.

Note that every lsRE fragment is ensured in the regex decomposition procedure to have at least one ldRE fragment with high pre-filtering effectiveness \footnote{Assuming random traffic, if the matching probability of one ldRE is lower than $P_T$, the ldRE is thought with high pre-filtering effectiveness. In our actual implementation, $P_T$ is set to 0.0001.}. Otherwise, the lsRE fragment is denoted as an unfriendly lsRE fragment, which is merged into neighboring lusRE fragments as described in Section \ref{sec:regex-splitting}.

After extracting one proper ldRE from each lsRE fragment, all the ldREs can then be employed to construct the xor filter as depicted in \cite{2020Xor}. Since each xor filter unit can only support the matching of patterns with the same length, multiple xor filter units are used to support different lengths as shown in Fig.\ref{fig:FPGA-CPU-arch}. The detailed implementation of the xor filter will be further discussed in Section \ref{sec:scheme-implementation}.

\subsection{Anchor DFA construction}
\label{sec:anchor-dfa-construction}

According to the ldRE position of each lsRE, we divide each lsRE fragment into two parts, the front part, and the back part. The front part for one lsRE starts from the head and ends at the end of the ldRE, while the back part is the remaining part of the lsRE. For $lsRE_{1-1}$="user=[a-f0-9]\{32\}" in Fig.\ref{fig:compilation-overview}, the front part is "user=" and the back part is "[a-f0-9]\{32\}". While for $lsRE_{4-1}$="\textbackslash d\{1,6\}\textbackslash x00mic\textbackslash x7c", since the extracted ldRE (i.e., "\textbackslash x00mic\textbackslash x7c") ends at the end of $lsRE_{4-1}$, it has only the front part, which is $lsRE_{4-1}$ itself.

After dividing each lsRE fragment, all front parts of the lsRE fragment set are compiled into one anchor DFA. While the back part of each lsRE fragment is compiled into one anchor DFA. Note that when the xor filter reports a match at some position $pos$, to perform the matching of all front parts for the lsRE set, the anchor DFA should match the input from $pos$ reversely. Therefore, the front parts of all lsREs are reversed\footnote{A reversed version of one regex matches the reverse of original accepted texts. For example, the reversed version of $lsRE_{1-1}$="user=[a-f0-9]\{32\}" is "[a-f0-9]\{32\}=resu".} first and then compiled into one anchor DFA, which is denoted as \textbf{reverse DFA} in the remaining text. As for the matching of the back part for each lsRE, since the corresponding anchor DFA only needs to perform it forward from $pos + 1$ as typical DFAs, the anchor DFA for each back part is denoted as one \textbf{forward DFA} in this paper.

Consequently, the anchor DFA engine in XAV includes one reverse DFA and multiple forward DFAs. The matching procedure of the anchor DFA engine is as below. When the xor filter reports some match at a position $pos$, a matching thread of the reverse DFA is activated at $pos$ and matches the input reversely. The reverse DFA can determine whether the input matches some front part of one lsRE fragment. If the reverse DFA is matched successfully by the input, the anchor DFA engine will find if there is one back part corresponding to the matched front part. If the back part exists, the corresponding forward DFA for the back part is then used to match the input from $pos + 1$ forward. If there is no back part to be matched or the back part is successfully matched, the anchor DFA engine will report the lsRE matching result to the further verification engine.

\subsection{Verification engine construction}
The verification engine is for the semantic verification of original regexes, which needs to record lsRE fragment matches and perform the matching of the lusRE fragment set. 

Any regex $RE$ can be regarded as the form of $R_1S_1R_2S_2...S_nR_{n+1}$ (for $1\leq i\leq n$, $R_i$ represents a lusRE fragment and $S_i$ represents a lsRE fragment), where $R_1$ and $R_{n+1}$ may be empty. To match $RE$, one text should match $R_1$, $S_1$, ..., $S_n$ and $R_{n+1}$ in order. As described earlier, the matching of each lsRE fragment (i.e., $S_i$ in $RE$) is performed by the xor filter and the anchor DFA engine. When some lsRE fragment $S_k$ is reported to match the input text $T$ starting at position $pos\_s_k$ and ending at position $pos\_e_k$, the verification engine should perform the semantic verification of $RE$ as follows. 

\begin{enumerate}
\item If $k=$1, the verification engine will check if $R_1$ matches $T$ ending at $pos\_s_k-1$. If $R_1$ is matched, then the matching result of $S_k$ will be recorded.
\item If $k\geq$2, the verification engine will first check whether there is one matching record of $S_{k-1}$. If not, the verification engine will directly ignore the matching result of $S_k$. Otherwise, the verification engine will check if $R_k$ can be matched starting from $pos\_e_{k-1}+1$ and ending at $pos\_s_k-1$. If $R_k$ is successfully matched, the matching result of $S_k$ will be recorded by the verification engine.
\item If $k=n$, since $S_n$ is the last lsRE fragment of $RE$, the verification engine will further check if $R_{n+1}$ is matched by $T$ starting at $pos\_e_n+1$. If $R_{n+1}$ is matched successfully, it indicates that all fragments of $RE$ have been matched by $T$ continuously. In this case, $RE$ will be reported to match the input text $T$.
\end{enumerate}

As for how to perform the matching of lusRE fragments, three matching approaches are employed for three kinds of lusREs, which are listed below.

\begin{itemize}
    \item \textbf{dot-MN} (".\{$m$,$n$\}"): The first kind of lusREs are denoted as dot-MN, which means that it matches any character for $m$ to $n$ times ($0\leq m\leq n$ and $n$ can be infinity). To check the semantics of one dot-MN lusRE, the verification engine only needs to judge whether the interval of two lsRE fragment matching results is in $(m, n)$.
    \item \textbf{CC-MN} ("[$c_1c_2...c_k$]\{$m$,$n$\}"): The second kind of lusREs are denoted as CC-MN, which is a character class with counting constraints. The semantics of one CC-MN is to match a text with $m$ to $n$ characters in [$c_1c_2...c_k$], where $c_i$ ($1\leq i\leq k$) is a literal character and [$c_1c_2...c_k$] is a character set representing the character class. To check the semantics of one CC-MN, the verification engine first judges whether the interval of two lsRE fragment matching results is in $(m, n)$. Then, all characters between the two lsRE fragment matching results should be checked if in [$c_1c_2...c_k$].
    \item \textbf{complex lusRE}: The third kind of lusREs are denoted as complex lusREs, which include all other lusREs except the first two kinds of lusREs. Since the semantics of one complex lusRE is much more complex, each complex lusRE is compiled into a DFA, which is denoted as \textbf{lusDFA} in this paper, to perform the matching.
\end{itemize}

\section{Implementation via an FPGA-CPU architecture}
\label{sec:scheme-implementation}
Since the matching logic of the xor filter and the anchor DFA engine is simple and suitable for hardware implementation. To achieve high matching performance, we use FPGA to implement the xor filter and the anchor DFA engine. While the logic of the verification engine is much more complex, which is suitable for software implementation in the CPU. Therefore, an FPGA-CPU architecture is designed to implement XAV in this paper. Through exploiting the massively parallel logic of FPGA, a high concurrent matching throughput is expected. 

\subsection{Architecture overview}
\begin{figure}[tb]
    \centering
    \centerline{\includegraphics[width=0.5\textwidth,trim=0 450 485 0,clip]{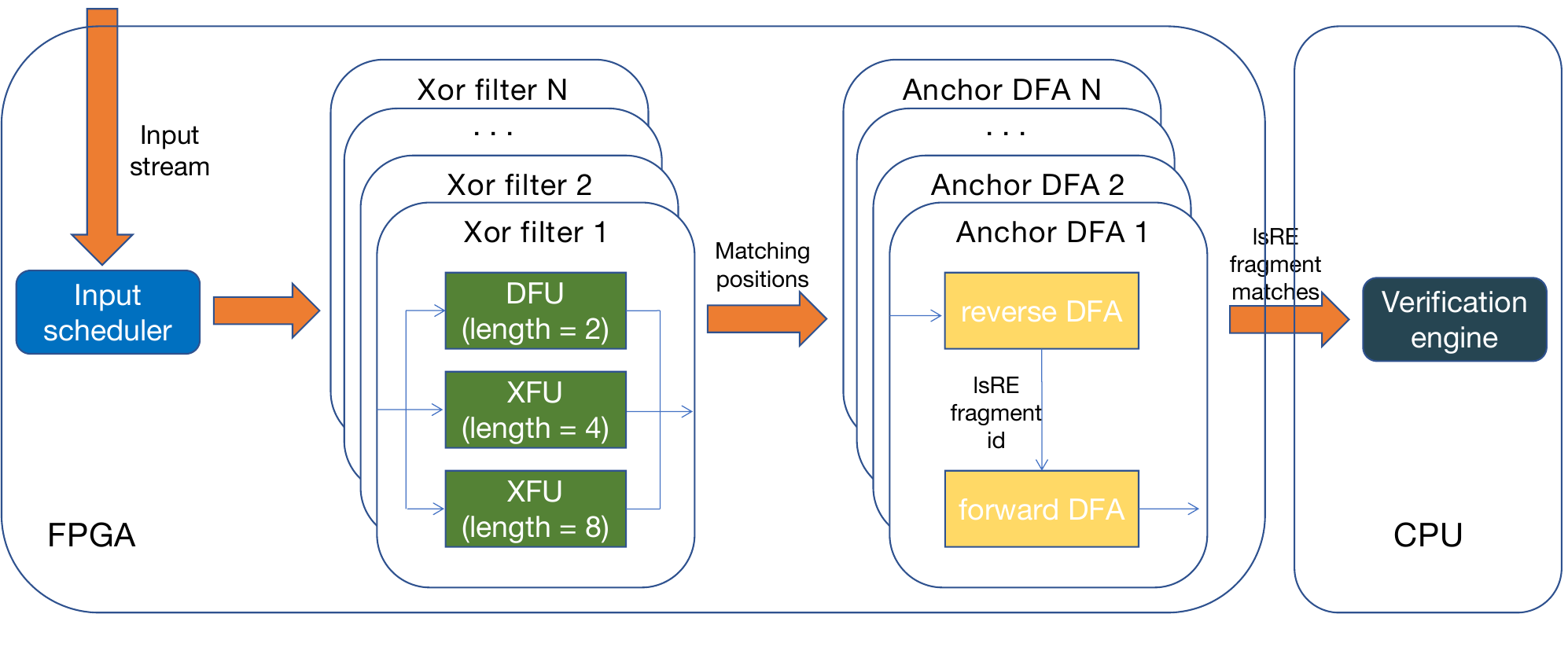}}
    \caption{FPGA-CPU architecture for XAV implementation.}
    \label{fig:FPGA-CPU-arch}
\end{figure}

The proposed FPGA-CPU architecture for XAV implementation is outlined in Fig.\ref{fig:FPGA-CPU-arch}. Firstly, the input packet stream is distributed to multiple \textbf{matching units}, where each matching unit comprises an xor filter and an anchor DFA engine. Each packet to be searched is then scanned by the xor filter in each matching unit. When some match occurs in the xor filter, the match position is reported to the anchor DFA in the same matching unit. The anchor DFA will start a matching thread at that position. If one lsRE match is found by the anchor DFA, the lsRE matching result is then reported to the CPU for further verification.

Note that when the CPU performs the verification of the semantics of CC-MN and complex lusREs, it needs to scan the packet content. Therefore, for those lsRE fragments that will cause CC-MN or complex lusRE semantic verification, once a match occurs, the packet needs to be transmitted to the CPU. When a large proportion of traffic is required to be processed in the CPU, the CPU overhead will be too heavy to make the overall REM scheme feasible. However, for practical applications, the practical traffic rarely matches one rule. Besides, when performing regex decomposition, the lsREs with low pre-filtering effectiveness have been merged into neighboring lusREs. Therefore, only a small proportion of traffic needs to be sent to the CPU in our scheme, and the CPU overhead is small, which will be shown in the evaluation section.

\subsection{Xor filter implementation}

In our detailed implementation, each xor filter comprises one direct filter unit (DFU) and two xor filter units (XFU) as shown in Fig.\ref{fig:FPGA-CPU-arch}. The direct filter unit is essentially a bitmap of size $2^{16}$, so it can be applied for matching ldREs with a length of 2. The two xor filter units are used for matching ldREs of length 4 and length 8 respectively. For ldREs with a length other than 2, 4, or 8, they will be cut to their nearest length in \{2, 4, 8\} in the compilation procedure. 

It is noted that the xor filter does not support the matching for ldREs of length 1. Since the matching probability for one ldRE of length 1 is at least $\frac{1}{256}$, which is too large to avoid the frequent activation of the anchor DFA engine.

\subsection{Anchor DFA implementation}
\label{sec:anchor-dfa-implementation}

\begin{figure}[tb]
    \centering
    \centerline{\includegraphics[width=0.5\textwidth,trim=0 140 505 0,clip]{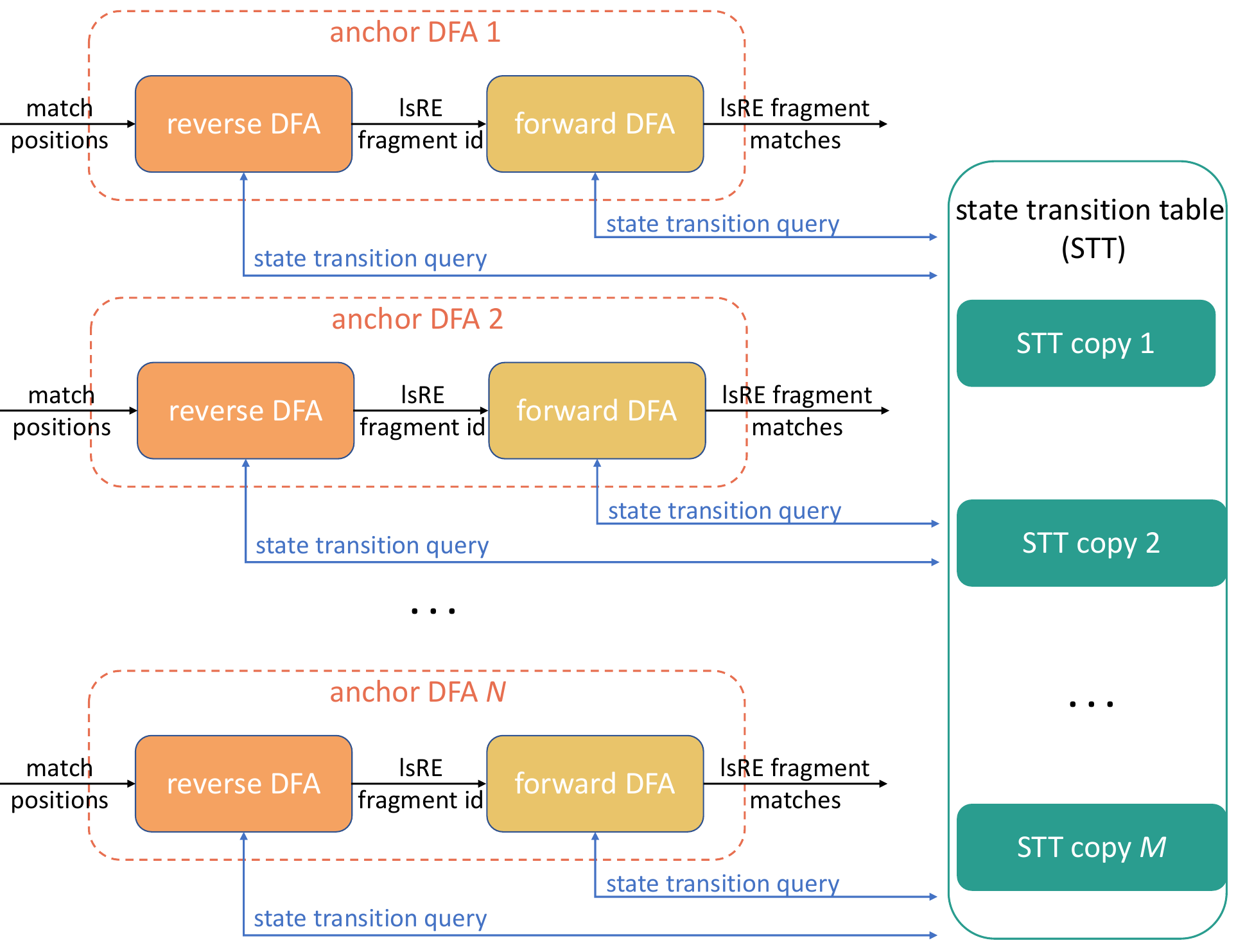}}
    \caption{All anchor DFAs share $M$ state transition tables.}
    \label{fig:STT-share}
\end{figure}

In each matching unit, the anchor DFA engine comprises a reverse DFA and a forward DFA. It is noticed that multiple forward DFAs generated in the compilation stage can be regarded as a forward DFA by merging the state transition tables when performing matching. Our implementation of the anchor DFA engine is as shown in Fig.\ref{fig:STT-share}.

For a large-scale rule-set containing thousands of regexes, the state transition table (STT) of the anchor DFA can consume a lot of memory. However, the memory resource is limited, and allocating one copy of STT for the anchor DFA in each matching unit is infeasible. Fortunately, the anchor DFA in each matching unit is active only for a small part of the time. Hence, the $N$ anchor DFAs in $N$ matching units can share a total of $M$ copies of STT ($M\leq N$) \footnote{In our detailed implementation, $N$ is 64 and $M$ is set to 6.} as shown in Fig.\ref{fig:STT-share}. Each state transition query emitted from all anchor DFAs (including the forward DFAs and the reverse DFAs in all matching units) is transmitted to a query scheduler, which will finish the query with an idle STT copy and return the query result to the corresponding anchor DFA.

Through state transition table sharing, the number of STT copies can be greatly reduced. However, for rule-sets containing thousands of regexes, if no compression techniques are employed, a few STT copies will still occupy up to hundreds of megabytes of memory, which far exceeds the internal storage of current FPGAs or the storage of fast external SRAM. Therefore, compressing the STT of the anchor DFA is necessary. In our implementation, a hybrid STT compression scheme, which combines the perfect hashing method \cite{2017Perfect} and the bitmap encoding method \cite{2004Deterministic}, is utilized. Experiments show that the hybrid compression scheme can effectively reduce the memory cost of anchor DFA by up to 98\%.

\section{Evaluation}
\label{sec:evaluation}
In this section, the proposed REM scheme, XAV, is comprehensively evaluated with both real and synthetic traffic and rule-sets.

\subsection{Experimental settings}
\subsubsection{Platform}

\begin{figure}[tb]
    \centering
    \centerline{\includegraphics[width=0.35\textwidth,trim=0 245 585 0,clip]{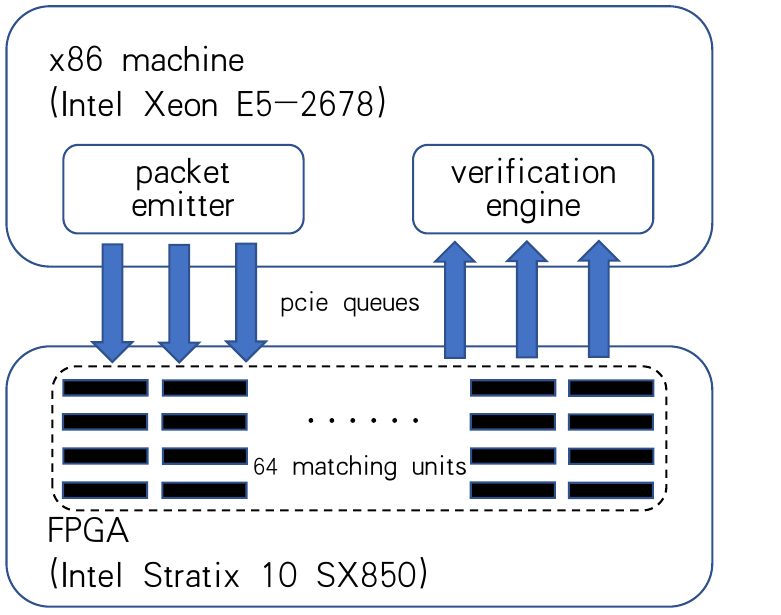}}
    \caption{The experimental environment which evaluates the performance of XAV.}
    \label{fig:platform}
\end{figure}

The experimental environment that evaluates the performance of XAV is shown in Fig.\ref{fig:platform}. The compiler and the verification engine of XAV are implemented with C++ 11. Experiments are carried out using an x86 machine with 128 GB DRAM. The CPU of the machine is a 2.5 GHz Intel Xeon E5-2678 and the operating system is CentOS 7.5. The xor filter and anchor DFA matching are performed with Intel Stratix 10 SX850 FPGA. In our implementation, the FPGA frequency is 200 MHz. As for the FPGA resource utilization, 96\% (807K/841K) logic elements and 75\% (6519KB/8692KB) memory bits are utilized.

\subsubsection{Rule-sets}
Nine regex rule-sets from Regex \cite{2010Regex} and ANMLZoo \cite{2016ANMLzoo} benchmark suits are used in our evaluation. The nine regex rule-sets show different characteristics, which are shown in Table \ref{tab:ruleset-character}. The scale and complexity of each rule-set are represented by the number of compiled NFA states and NFA transitions. From the table, the scale and complexity of rule-sets increase from top to bottom. \textit{Bro217} rule-set is the smallest and simplest one, while \textit{Snort} rule-set is the largest and most complex one.

\begin{table}[!tb]
\centering
\caption{The characteristics of test rule-sets.}
\label{tab:ruleset-character}
\begin{tabular}{ccccc}
\hline
Rule-set  & \#Rules & Avg. length & \#NFA states & \#NFA transitions \\ \hline
Bro217    & 217     & 12.33               & 2864         & 6583              \\
Ranges1   & 299     & 41.99               & 13338        & 19474             \\
Ranges05  & 300     & 42.19               & 13261        & 18080             \\
Dotstar03 & 300     & 40.29               & 13049        & 41977             \\
Dotstar06 & 300     & 41.91               & 13721        & 65447             \\
Dotstar09 & 299     & 40.98               & 13580        & 88103             \\
PowerEN   & 2860    & 17.70               & 47492        & 239344            \\
ClamAV    & 515     & 96.07               & 52040        & 1435257           \\
Snort     & 3170    & 62.24               & 212731       & 1586731           \\ \hline
\end{tabular}
\end{table}

\subsubsection{Traffic}
Two types of network traffic are used in evaluating the performance of XAV. The first type of traffic is 13.7 MB packets randomly generated and labeled as \textit{Random}. The second type of traffic is realistic packets captured in a campus network \footnote{The realistic network traffic was collected from volunteers, and the use of it has been approved by them.}. Realistic traffic is divided into three parts and they are labeled as \textit{Reality-1}, \textit{Reality-2}, and \textit{Reality-3}, respectively. The sizes of the three parts of realistic traffic are 160.5 MB, 70.2 MB, and 54.7 MB, respectively.

\subsection{Compilation time}
Table \ref{tab:ruleset-compilation-time} shows the compilation time of XAV on test rule-sets. The compilation procedure of XAV comprises four stages: regex decomposition, xor filter compilation, anchor DFA compilation, and verification engine compilation. As shown in the table, XAV compiles rapidly on test rule-sets. For the first six small rule-sets, their total compilation time is almost less than 1 second. For the two medium-scale rule-sets, \textit{PowerEN} and \textit{ClamAV}, the compilation of XAV costs a total of only several seconds. For the largest rule-set, \textit{Snort}, XAV can also finish the entire compilation in about 30 seconds.

\begin{table*}[t]
\centering
\caption{The compilation time (seconds) of XAV on test rule-sets.}
\label{tab:ruleset-compilation-time}
\begin{tabular}{cccccc}
\hline
Rule-set  & \begin{tabular}[c]{@{}c@{}}Regex\\ decomposition\end{tabular} & \begin{tabular}[c]{@{}c@{}}Xor filter\\ compilation\end{tabular} & \begin{tabular}[c]{@{}c@{}}Anchor DFA\\ compilation\end{tabular} & \begin{tabular}[c]{@{}c@{}}Verification engine\\ compilation\end{tabular} & Total  \\ \hline
Bro217    & 0.006                                                         & 0.003                                                            & 0.159                                                            & 7.13x10$^{-5}$                                                                 & 0.168  \\
Ranges1   & 0.024                                                         & 0.001                                                            & 0.957                                                            & 1.09x10$^{-4}$                                                                 & 0.982  \\
Ranges05  & 0.025                                                         & 0.001                                                            & 0.976                                                            & 9.75x10$^{-5}$                                                                 & 1.002  \\
Dotstar03 & 0.024                                                         & 0.002                                                            & 0.940                                                            & 1.57x10$^{-4}$                                                                 & 0.965  \\
Dotstar06 & 0.025                                                         & 0.002                                                            & 0.929                                                            & 2.12x10$^{-4}$                                                                 & 0.956  \\
Dotstar09 & 0.026                                                         & 0.002                                                            & 0.936                                                            & 3.51x10$^{-4}$                                                                 & 0.964  \\
PowerEN   & 0.143                                                         & 0.019                                                            & 4.750                                                            & 0.003                                                                     & 4.914  \\
ClamAV    & 0.102                                                         & 0.004                                                            & 4.879                                                            & 0.189                                                                     & 5.174  \\
Snort     & 1.293                                                         & 0.050                                                            & 27.104                                                           & 3.652                                                                     & 32.100 \\ \hline
\end{tabular}
\end{table*}

Among the four XAV compilation stages, anchor DFA compilation costs most of the overall compilation time. The reason is that anchor DFA compilation involves the DFA construction process \cite{zhong2019accelerating}, which runs relatively slowly. Although verification engine compilation can also involve DFA construction for generating lusDFAs (DFAs for complex lusREs), it usually takes much less time for that the overall size of all lusDFAs is much smaller than the size of anchor DFA.

\subsection{Spatial performance}
Table \ref{tab:spatial-performance} lists the spatial data of XAV on test rule-sets.

\begin{table*}[tb]
\centering
\caption{The spatial performance of XAV on test rule-sets.}
\label{tab:spatial-performance}
\begin{tabular}{ccccccc}
\hline
\multirow{2}{*}{Rule-set} & Xor filter                                                        & \multicolumn{3}{c}{Anchor DFA}                                                                                                                                                           & \multicolumn{2}{c}{Verification engine}                                                                                           \\ \cline{2-7} 
                          & \begin{tabular}[c]{@{}c@{}}memory\\ consumption (KB)\end{tabular} & \begin{tabular}[c]{@{}c@{}}state\\ number\end{tabular} & \begin{tabular}[c]{@{}c@{}}memory\\ consumption (KB)\end{tabular} & \begin{tabular}[c]{@{}c@{}}compression\\ ratio\end{tabular} & \begin{tabular}[c]{@{}c@{}}lusDFA\\ state number\end{tabular} & \begin{tabular}[c]{@{}c@{}}memory\\ consumption (KB)\end{tabular} \\ \hline
Bro217                    & 18.850                                                            & 2302                                                   & 24.775                                                            & 0.016                                                       & 0                                                             & 1.680                                                             \\
Ranges1                   & 8.405                                                             & 10759                                                  & 95.850                                                            & 0.014                                                       & 0                                                             & 2.570                                                             \\
Ranges05                  & 8.404                                                             & 10724                                                  & 88.240                                                            & 0.013                                                       & 0                                                             & 2.578                                                             \\
Dotstar03                 & 8.506                                                             & 9804                                                   & 88.220                                                            & 0.014                                                       & 0                                                             & 3.273                                                             \\
Dotstar06                 & 8.586                                                             & 9692                                                   & 87.261                                                            & 0.014                                                       & 0                                                             & 3.969                                                             \\
Dotstar09                 & 8.647                                                             & 8900                                                   & 81.614                                                            & 0.014                                                       & 0                                                             & 4.656                                                             \\
PowerEN                   & 14.435                                                            & 38698                                                  & 299.060                                                           & 0.011                                                       & 0                                                             & 42.508                                                            \\
ClamAV                    & 9.194                                                             & 38078                                                  & 431.008                                                           & 0.018                                                       & 761                                                           & 768.555                                                           \\
Snort                     & 12.646                                                            & 135296                                                 & 1601.953                                                          & 0.019                                                       & 35306                                                         & 35447.742                                                         \\ \hline
\end{tabular}
\end{table*}

The second column of the table shows the xor filter memory consumption of XAV. Since each entry in the xor filter consumes an average of only a few bits, the memory consumption of the xor filter on each test rule-set is small and it varies from a few kilobytes to a dozen kilobytes.

The spatial data related to anchor DFA are listed in the third to fifth columns of Table \ref{tab:spatial-performance}. As anchor DFA has simplified regex semantics, it can avoid the DFA state explosion problem as stated in Section \ref{sec:motivation-anchor-dfa}. Consequently, the anchor DFA state number of each rule-set is close to the NFA state number listed in Table \ref{tab:ruleset-character}. After state table compression, the anchor DFA memory consumption for the first six small rule-sets is dozens of kilobytes. While for the two medium-scale rule-set, \textit{PowerEN} and \textit{ClamAV}, the anchor DFA occupies a few hundred kilobytes of memory. For \textit{Snort}, the largest test rule-set, the anchor DFA memory consumption can reach 1600 kilobytes. The compression ratio of the anchor DFA state table is from 0.011 to 0.019 for each rule-set, which means that state table compression helps reduce the memory consumption of anchor DFA by 98\% to 99\%.

Note that as the anchor DFA in each matching unit is active for only a small part of the time, multiple matching units can share one copy of anchor DFA STT (state transition table) as explained in Section \ref{sec:anchor-dfa-implementation}. In the implementation, an average of about every ten matching units is designed to share one anchor DFA STT. Therefore, the average anchor DFA memory consumption per matching unit is about $\frac{1}{10}$ that is listed in Table \ref{tab:spatial-performance}.

The spatial data related to the verification engine are listed in the sixth and seventh columns of Table \ref{tab:spatial-performance}. For all small test rule-sets and \textit{PowerEN} rule-set, as no complex lusREs are generated in the regex decomposition stage, the state number of lusDFAs is zero and the memory consumption of the verification engine is small. When no lusDFAs are generated, the verification engine only mainly involves one mapping table recording the connection relations of each lsRE. Consequently, the memory consumption of the verification engine is small and only a few kilobytes of memory are required. For \textit{ClamAV} and \textit{Snort} rule-sets, as complex lusREs result in the generation of lusDFAs, the verification engine consumes much more memory. In order to perform fast verification within the CPU, no compression techniques are applied for lusDFA STT and each lusDFA state requires one kilobyte of memory. As \textit{Snort} rule-set generates tens of thousands of lusDFA states, it requires up to tens of MB of memory for the verification engine. Compared with the memory consumption of at least hundreds of MB in most existing DFA-based REM schemes, XAV is believed to have shown excellent spatial performance.

\subsection{Temporal performance}
\subsubsection{Effectiveness of xor filter}

\begin{figure}[tb]
    \centering
    \centerline{\includegraphics[width=0.5\textwidth,trim=0 20 70 20,clip]{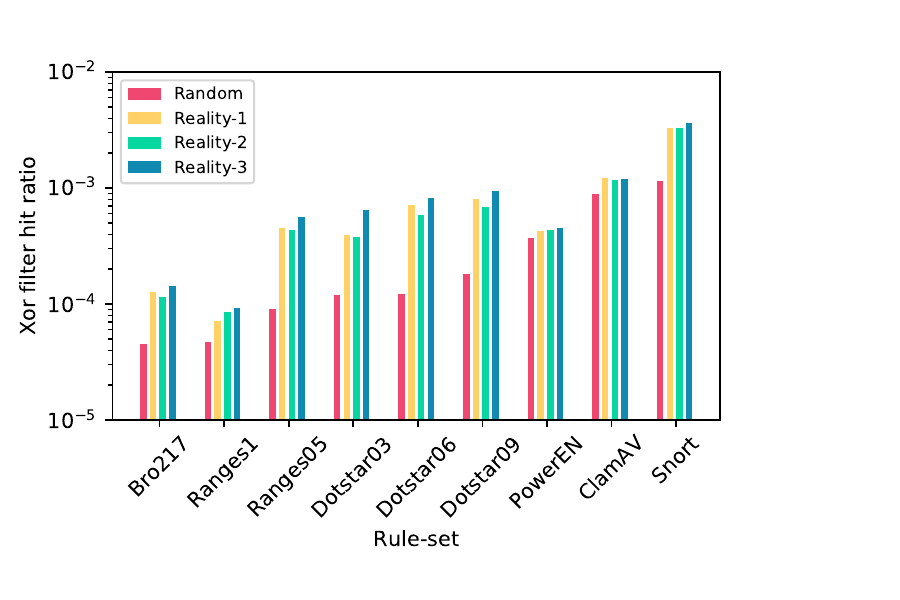}}
    \caption{Xor filter hit ratio for each rule-set and each type of traffic.}
    \label{fig:xor-filter-ratio}
\end{figure}

The xor filter hit ratio for each rule-set and each type of traffic is shown in Figure \ref{fig:xor-filter-ratio}.

It can be seen from the figure that the xor filter has high prefiltering efficiency. The xor filter hit ratio is lower than 0.005 on all test rule-sets and all test traffic, which means that the xor filter helps reduce the number of anchor DFA matching threads by more than 99.5\%. The figure also demonstrates that the xor filter hit ratio rises with the increase of rule-set scale and complexity. It is reasonable for that complex and large-scale rule-sets generate more lsRE fragments and therefore more entries for the xor filter to match. It is also noted that the filtering effectiveness of the xor filter on random traffic is obviously higher than that on reality traffic.

\subsubsection{Anchor DFA matching overhead}
\begin{figure}[tb]
    \centering
    \centerline{\includegraphics[width=0.5\textwidth,trim=0 20 70 20,clip]{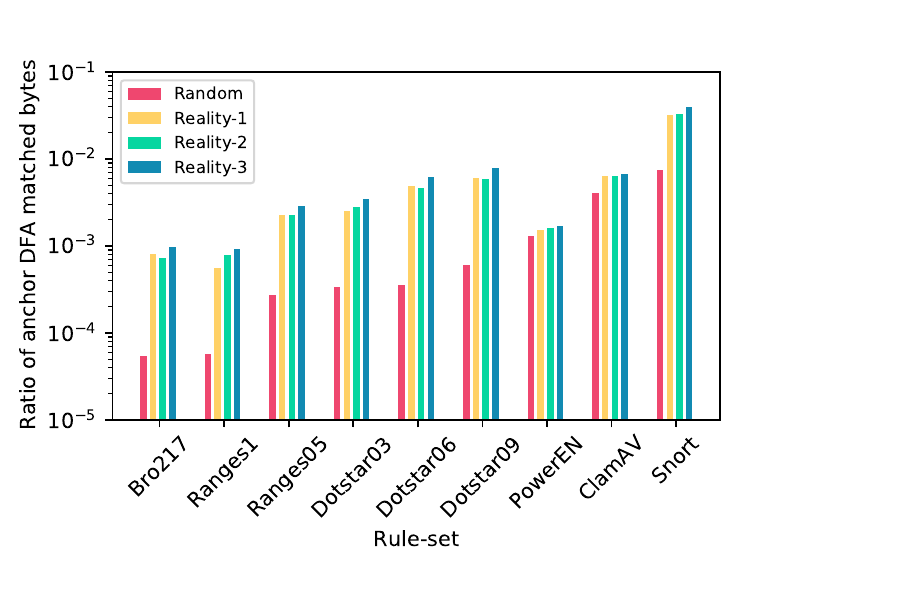}}
    \caption{Ratio of anchor DFA matched bytes for each rule-set and each type of traffic.}
    \label{fig:adfa-match-ratio}
\end{figure}

The anchor DFA matching overhead in XAV is represented by the ratio of the matching bytes of the anchor DFA to the total traffic. As shown in Figure \ref{fig:adfa-match-ratio}, the ratio of anchor DFA matched bytes on each rule-set and each type of traffic is lower than 0.05, which is a quite small value and indicates that every 20 bytes of traffic can trigger at most one state transition for the anchor DFA. Consequently, as mentioned in Section \ref{sec:anchor-dfa-implementation}, about every ten matching units can share one copy of the anchor DFA STT, which is enough to make the anchor DFA not the performance bottleneck of XAV.

\subsubsection{Verification engine overhead}
\begin{figure}[tb]
    \centering
    \centerline{\includegraphics[width=0.5\textwidth,trim=0 20 70 20,clip]{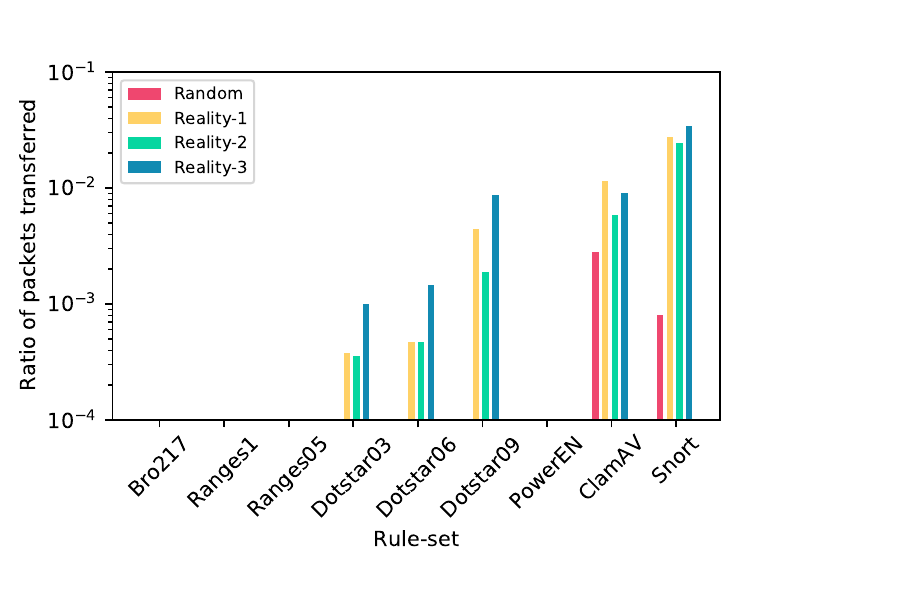}}
    \caption{Ratio of packets transferred from FPGA to CPU for each rule-set and each type of traffic.}
    \label{fig:transfer-packets-ratio}
\end{figure}

\begin{figure}[tb]
    \centering
    \centerline{\includegraphics[width=0.5\textwidth,trim=0 20 70 20,clip]{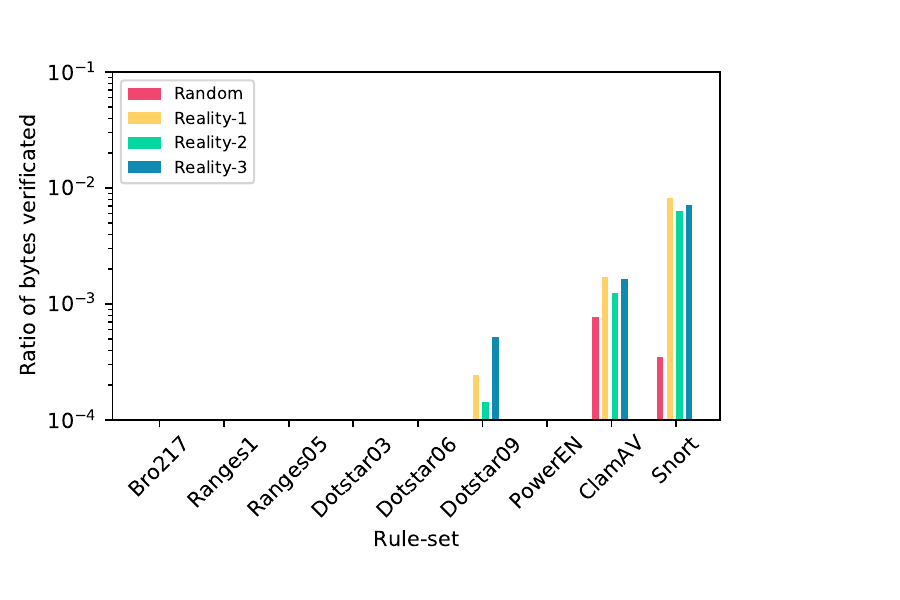}}
    \caption{Ratio of bytes needing to be matched by the verification engine for each rule-set and each type of traffic.}
    \label{fig:ve-bytes-ratio}
\end{figure}

The burden of the verification engine on the CPU comes from two aspects: obtaining partial traffic packets from FPGA and matching partial traffic bytes. Figure \ref{fig:transfer-packets-ratio} shows the ratio of packets needing to be transferred from FPGA to CPU on each rule-set and each type of traffic, and Figure \ref{fig:ve-bytes-ratio} shows the ratio of traffic bytes needing to be matched by the verification engine.

From the two figures, for rule-sets \textit{Bro217}, \textit{Ranges1}, \textit{Ranges05}, and \textit{PowerEN}, the ratios of packets transferred and bytes matched by the verification engine are both zero, which indicates there is no verification overhead. The zero verification overhead for the four rule-sets is because each regex in these rule-sets is a lsRE itself and does not need to be decomposed. Therefore, the anchor DFA in XAV can perform the entire matching of each regex in these rule-sets and no verification is required.

For the other five rule-sets, the ratio of bytes transferred from FPGA to CPU and the ratio of bytes matched by the verification engine are all small. Even for the largest \textit{Snort} rule-set and realistic traffic, the ratio of bytes transferred and the ratio of bytes matched by verification are lower than 0.04 and 0.01 respectively. To make the verification engine not the performance bottleneck of XAV, a few CPU cores are enough for the small verification overhead. In the evaluation of XAV, one to six CPU cores are sufficient to reach the maximum matching throughput for all rule-sets.

\subsection{Comparison with existing REM schemes}

\subsubsection{Comparison with commodity schemes}

\begin{figure}[tb]
    \centering
    \centerline{\includegraphics[width=0.5\textwidth,trim=10 20 30 20,clip]{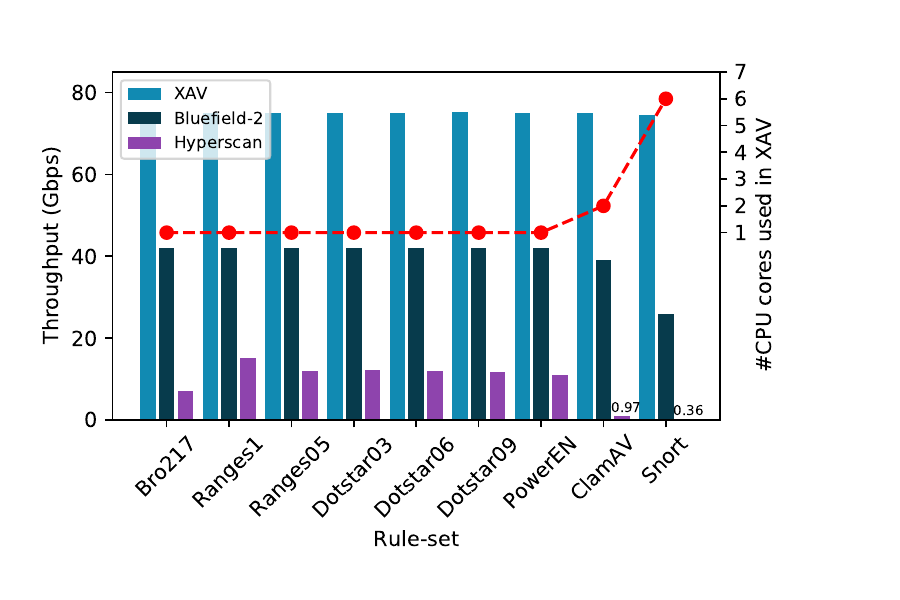}}
    \caption{Throughput comparison with commodity REM schemes on realistic traffic.}
    \label{fig:throughput-compare}
\end{figure}

In our evaluation, the regex matching performance of XAV is compared with that of two commodity schemes, Bluefield-2 \cite{Burstein2021NvidiaDC} and Hyperscan \cite{wang2019hyperscan}. Bluefield-2 is a data processing unit (DPU) launched by Nvidia, which can provide a variety of acceleration functions, including accelerated regular expression matching. Hyperscan is the current most efficient software REM solution provided by Intel. The throughput comparison is shown in Figure \ref{fig:throughput-compare}.

In our implementation, with 64 matching units, XAV can achieve a maximum throughput of about 75 Gbps\footnote{The performance bottleneck lies in the transmission of packets through PCIE channels.
} on all rule-sets. In comparison, Bluefield-2 achieves 42 Gbps throughput for most rule-sets, but suffers performance degradation for \textit{Snort} rule-set, on which the throughput is 26 Gbps. What is worse, hundreds of rules in \textit{Snort} rule-set are not supported by the compiler of Bluefield-2. Although XAV has to extract string fragments from each regex rule to build the xor filter, only tens of \textit{Snort} rules containing no string fragment are not supported by XAV. Therefore, it can be drawn that XAV achieves better matching throughput and supports more complex regex rules than Bluefield-2.

For small rule-sets, Hyperscan can achieve a matching throughput of several to even more than ten Gbps with a CPU core, which is an excellent performance for software REM schemes. However, with the rule-set complexity and scale increase, the matching throughput of Hyperscan drops significantly and is only 0.97 Gbps on \textit{ClamAV} rule-set and 0.36 Gbps on \textit{Snort} rule-set. As a software solution, the biggest advantage of Hyperscan is that it can support nearly all \textit{Snort} rules for software-only agility. It is noted that in the matching throughput comparison on \textit{Snort} rule-set, Hyperscan only compiles the \textit{Snort} rules supported by XAV for fairness. If Hyperscan compiles all \textit{Snort} rules, the matching performance of Hyperscan will downgrade remarkably to only tens of Mbps. Although XAV is less flexible than Hyperscan according to supporting as many complex regex rules as possible, XAV achieves much better regex matching performance than Hyperscan and therefore can help save a lot of CPU resources for kinds of network security applications. For small rule-sets, XAV can reduce the requirement of more than 80\% CPU cores compared to Hyperscan. While for large rule-sets like \textit{ClamAV} and \textit{Snort}, XAV helps to offload more than 95\% computation burden into FPGA.

\subsubsection{Comparison with FPGA-based schemes}

Table \ref{tab:comparison-FPGA} compares XAV to other FPGA-based schemes. The data for other FPGA-based schemes are gained from their papers. From the table, it shows that XAV has a significantly shorter compilation time than other FPGA-based schemes. The throughput efficiency metric shows that XAV provides the highest matching performance under the same hardware resource consumption. Compared to other FPGA-based schemes, XAV achieves throughput efficiency ranging from 2.5 to 19 times higher, according to the matching performance achieved per unit of logic/storage resources.

\begin{table*}[t]
\centering
\caption{Comparison with FPGA-based schemes on \textit{Snort} rule-set.}
\label{tab:comparison-FPGA}
\begin{tabular}{ccccccc}
\hline
Scheme       & Platform         & \begin{tabular}[c]{@{}c@{}}Rule-set\\ size\end{tabular} & \begin{tabular}[c]{@{}c@{}}Compilation\\ time\end{tabular} & \begin{tabular}[c]{@{}c@{}}Throughput\\ (Gbps)\end{tabular} & \begin{tabular}[c]{@{}c@{}}Tput efficiency\\ ($\frac{Gbps*states}{\#LUT}$)\end{tabular} & \begin{tabular}[c]{@{}c@{}}Tput efficiency\\ ($\frac{Gbps*states}{bytes}$)\end{tabular} \\ \hline
Xie et al. \cite{10.1145/3314576,2017REAPR}  & Xilinx Kintex    & 3379                                                    & $\sim$10 hours                                             & 4.16                                                        & $\sim$4.16                                                                        & $\sim$0.21                                                                        \\
Yant et al. \cite{2012High} & Xilinx Virtex-5  & 565                                                     & hours scale                                                & 10.30                                                       & 10.1                                                                              & \textgreater{}0.77                                                                \\
Yant et al. \cite{MDFA-yang2018high} & Xilinx Virtex-7  & 34                                                     & N/A                                                & 141                                                       & $\sim$0.43                                                                              & $\sim$0.34                                                                \\
Zhong et al. \cite{10.1093/comjnl/bxac138} & Intel Arria 10   & 3420                                                      & $\sim$7 minutes                                            & 6.33                                                        & 2.97                                                                              & 0.23                                                                              \\
XAV          & Intel Stratix 10 & 3170                                                    & 32 seconds                                                 & 75                                                          & 56.58                                                                         & 1.89                                                                         \\ \hline
\end{tabular}
\end{table*}

The first two FPGA schemes in Table \ref{tab:comparison-FPGA} are based on NFA. Since NFA needs to be mapped into specific FPGA logic for regex matching and the synthesis of FPGA circuits is quite time-consuming, the compilation time of the two FPGA NFA schemes is lengthy and can take several hours. The next two FPGA schemes and XAV are all based on DFA. As DFA can be updated by rewriting the internal memory in FPGA, the compilation of the latter three schemes can avoid the time-consuming FPGA circuit resynthesis. Consequently, the compilation time for rule-set updates of the latter three FPGA DFA schemes is significantly shorter than the first two FPGA NFA schemes. The third FPGA scheme \cite{MDFA-yang2018high} does not solve the state explosion problem of DFA, resulting in that it only supports small rule-sets of tens of regexes. Compared to the fourth FPGA scheme, because XAV better avoids the state explosion problem of DFA by utilizing anchor DFA, XAV generates much smaller DFA than the scheme proposed in \cite{10.1093/comjnl/bxac138}. As DFA compilation is also a time-consuming task\cite{zhong2019accelerating}, XAV takes much less time to complete the construction of smaller DFA.

Compared to other FPGA schemes, the much higher matching throughput efficiency achieved by XAV is mainly due to the use of the xor filter. Because each xor filter unit occupies very little memory and can prevent the majority of traffic from being matched by the latter stages, tens of xor filter units can be implemented in FPGA. Consequently, high matching throughput efficiency can be achieved in XAV.

\section{Conclusion}
\label{sec:conclusion}
Regular expression matching plays a crucial role in various network security applications. High-performance regex matching has always been desired to help implement real-time packet checking. This paper proposes to use anchor DFA to tackle the state explosion problem of DFA, thereby supporting large-scale regex rule-sets. To achieve high matching performance, the xor filter and an extra verification stage are introduced to cooperate with anchor DFA. Consequently, the proposed scheme is named XAV according to its three matching stages: xor filter, anchor DFA, and verification. Since the xor filter and anchor DFA have simple logic and are suitable for hardware implementation, an FPGA-CPU architecture is employed to implement XAV for high matching throughput. Our evaluation shows that XAV has fantastical space and temporal complexity. Compared to Bluefield-2 and Hyperscan, the FPGA-CPU implementation of XAV shows a much higher matching throughput of up to 75 Gbps for the large \textit{Snort} rule-set. Compared to other FPGA-based schemes, XAV not only performs better in matching throughput but also consumes fewer hardware resources.

\bibliographystyle{unsrt}
\bibliography{\jobname}

{\appendix[Preliminaries]
\label{sec:preliminary}
In this section, the background knowledge about regular expressions and finite state automata is introduced.

\subsection{Regular expressions}

A regex is a sequence of characters that define a search pattern, namely languages or a string set. Each character in a regex is either a general character or a metacharacter. Each metacharacter has a special meaning. For example, metacharacter \textit{“.”} means to match any character, and metacharacter \textit{“*”} means to match a character any number of times. Table \ref{tab:metacharacter} lists some common metacharacters used in regexes.

\begin{table*}[htbp]
  \small
  \centering
  \caption{The common metacharacters in regexes}
  \label{tab:metacharacter}
  \begin{tabular}{ccc}
  \hline
  Metacharacter & Semantics                                             & Example                                    \\ \hline
  .             & any single character                                 & “a.b” matches “aab”, “abb” and “acb”, etc  \\ 
  {[}{]}        & any single character in square brackets              & “{[}ab{]}” matches “a” or “b”              \\ 
  {[}\textasciicircum{}{]} & any single character not in square brackets       & “{[}\textasciicircum{}ab{]}” matches any character other than “a” and “b” \\ 
  *             & match the previous character any number of times     & “a*b” matches “b”, “ab”, and “aab”, etc    \\ 
  +             & match the previous character once or more times      & “a+b” matches “ab”, “aab”,  etc            \\ 
  \{m,\}                   & the previous element is repeated at least m times & “a\{2,\}” matches “aa”, “aaa”, and “aaaa”, etc                            \\ 
  \{m,n\}       & the previous element is repeated m to n times        & “a\{2,4\}” matches “aa”, “aaa”, and “aaaa” \\ 
  $\vert$             & match the preceding element or the following element & “a$\vert$b” matches “a” or “b”                   \\ 
  \textasciicircum{}       & match the start position of one string            & “\textasciicircum{}a” matches the strings start with “a”                  \\ 
  \$            & match the end position of one string                 & “a\$” matches the strings end with “a”     \\ \hline
  \end{tabular}
\end{table*}

\subsection{Finite state automata}

Regexes are usually first compiled into a finite state automaton for matching. It is theoretically proved that regexes and finite state automata are mathematically equivalent \cite{Hopcroft1979Introduction}. For any regex $r$, there is always a finite automaton $\mathcal{A}$ whose semantics are the same as $r$; on the contrary, for any finite automaton $\mathcal{A}$, there is always a regex $r$ whose semantics are the same as $\mathcal{A}$. 

Nondeterministic Finite Automaton (NFA) and Deterministic Finite Automaton (DFA) are the two most common automata applied for regex matching. Using the classical Thompson algorithm \cite{thompson1968programming}, a set of regexes can be compiled into an NFA. Furthermore, through the subset construction algorithm \cite{Hopcroft1979Introduction}, an NFA can be converted to a DFA. To eliminate redundant states, one DFA is usually minimized using the minimization algorithm \cite{DFA_mini}.

NFA and DFA can both be represented by a five-tuple ($\mathcal{Q}$, $\Sigma$, $\delta$, $q_0$, $\mathcal{F}$), which is interpreted as follows:
\begin{itemize}
    \item $\mathcal{Q}$: A finite set of states representing the state space of the automaton.

    \item $\Sigma$ : A finite set of input symbols (usually 256 ASCII characters).

    \item $\delta$ : The state transition function( $\mathcal{Q} \times \Sigma \to \mathcal{Q}$), indicating how the state transitions are performed for each input character.

    \item $q_0$: An initial state representing the state in which the automaton starts to match. ($q_0 \in \mathcal{Q}$)

    \item $\mathcal{F}$ : The set of matching/accepting states. It indicates that rule matching occurs when jumping to these states in the automaton. ($\mathcal{F} \subseteq \mathcal{Q}$)
\end{itemize}

Taking one DFA as an example to clarify the matching process of one automaton. Let the input text to be matched be $c_0c_1c_2...c_p$. To begin with, the DFA starts from the initial state $q_0$, which is called the current active state. For the first character $c_0$, the state transition function $\delta$ is invoked to get the next active state, which is supposed as $q_1=\delta(q_0,c_0)$. Then, the DFA reads the next character $c_1$, and consults the next active state, $q_2=\delta(q_1,c_0)$. The process is repeated until the end of the input text. Assume that the DFA visits a sequence of states, $q_0q_1...q_p$ during the whole matching process. If $q_p$ is in the accepting state set $\mathcal{F}$, it indicates that the DFA accepts the input text. Otherwise, the input text is rejected by the DFA.

The matching procedure of NFA is similar to DFA. The only difference lies in that in the matching process of one NFA, there can be multiple current active states, and each state can transit to multiple states through the state transition function $\delta$.

\subsection{NFA vs DFA}

Table \ref{tab:NFAvsDFA} lists the comparison between NFA and DFA according to the worst-case time complexity and space complexity. In general, employing either NFA or DFA is difficult to realize efficient regex matching for large-scale and complex rule-sets.

\begin{table}[tb]
\centering
\caption{Comparison between NFA and DFA}
\label{tab:NFAvsDFA}
\begin{tabular}{ccc}
\hline
Automaton    & Time complexity & Space complexity \\ \hline
NFA & $O(mn^2)$       & $O(mn)$          \\
DFA & $O(1)$          & $O(2^{mn})$      \\ \hline
\end{tabular}
\end{table}

NFA usually has a simpler structure than DFA. Theoretical analysis has shown that for a regex with length $n$,  NFA has a linear space complexity of $O(n)$. However, multiple states may be activated simultaneously in NFA. In the worst case, all states in NFA are active, and each state may jump to all other states. Therefore, for a regex of length $n$, the worst-case time complexity for matching each input character is $O(n^2)$. Compiling $m$ regexes into an NFA, the space complexity increases linearly to $O(mn)$, and the worst-case time complexity rises up to $O(mn^2)$, which indicates a low matching performance.

In contrast, since there is always one active state in DFA and only one state transition is required per input character, DFA has a low time complexity of is $O(1)$. The low time complexity indicates that DFA has a huge advantage in matching performance compared to NFA. However, the excellent temporal performance of DFA comes at the cost of huge space consumption. The subset construction algorithm actually traverses all possible state combinations in NFA and transforms each state combination into a new state in DFA. Therefore, for a regex of length $n$, the worst-case space complexity of DFA is as high as $O(2^n)$. For $m$ regexes, the worst-case space complexity of DFA increases dramatically to $O(2^{mn})$. When compiling a large-scale and complex rule-set into a DFA, the required space usually exceeds the upper limit of the compiler's physical memory. In this case, the state explosion of DFA occurs.}

\end{document}